\documentclass{elsart}

\newcommand{\be}{\begin{equation}}
\newcommand{\ee}{\end{equation}}
\newcommand{\lb}{\label}
\newcommand{\ol}{\overline}
\newcommand{\ba}{{\bf a}}
\newcommand{\bF}{{\bf f}}
\newcommand{\br}{{\bf r}}
\newcommand{\bu}{{\bf u}}
\newcommand{\bx}{{\bf x}}
\newcommand{\cL}{{\mathcal L}}
\newcommand{\btau}{{\mbox{\boldmath $\tau$}}}
\newcommand{\bomega}{{\mbox{\boldmath $\omega$}}}
\newcommand{\grad}{{\mbox{\boldmath $\nabla$}}}
\newcommand{\bdot}{{\mbox{\boldmath $\cdot$}}}
\newcommand{\btimes}{{\mbox{\boldmath $\times$}}}

\usepackage{natbib}

 \usepackage{graphics}

\begin{document}

\begin{frontmatter}



\title{Turbulent Cascade of Circulations}


\author{Gregory L. Eyink}

\address{CNLS, Los Alamos National Laboratory,
Los Alamos, NM 87545 \\ {\it and}\\ Department of Applied  Mathematics
\& Statistics, The Johns Hopkins University, Baltimore, MD  21210}
\ead{eyink@ams.jhu.edu}

\begin{abstract}
The circulation around any closed loop is a Lagrangian invariant for classical,
smooth solutions of the incompressible Euler equations in any number of space
dimensions. However,  singular solutions relevant to turbulent flows need
not preserve the classical integrals of motion. Here we generalize the Kelvin
theorem on conservation of circulations to distributional solutions of Euler
and
give necessary conditions for the anomalous dissipation of circulations.  We
discuss the important role of Kelvin's theorem in turbulent vortex-stretching
dynamics
and conjecture a version of the theorem which may apply to suitable singular
solutions.

\end{abstract}

\begin{keyword}
turbulence \sep circulation  \sep Euler equations  \sep Kelvin theorem
\end{keyword}

\end{frontmatter}

\section{Introduction}
\label{intro}



In a monumental paper, \cite{Helmholtz:1858}  formulated the fundamental laws
of vortex
motion for incompressible fluids. These include the statements, in three space
dimensions,
that  vortex lines are material lines and that the flux within any vortex tube
is a Lagrangian
invariant. Lord \cite{Kelvin:1869} gave an elegant alternative formulation of
these laws in terms
of the conservation of circulation, for any closed loop advected by an ideal
fluid. This theorem
is equally valid in any space dimension. However,  it is only rigorously proved
for
sufficiently smooth solutions. As was pointed out by \cite{Onsager:1949}, the
classical conservation
laws need not be valid for singular solutions of Euler equations. In
particular, breakdown of
the energy conservation law can account for the anomalous dissipation of energy
observed
in turbulent fluids at high Reynolds numbers.  From a physical point of view,
this breakdown
of energy conservation corresponds to the turbulent energy cascade and a flux
of energy to
arbitrarily small scales.  See also \cite{Eyink:1994},\cite{Constantin et
al.:1994},
\cite{DuchonRobert:2000}.

These considerations make it very natural to inquire whether Kelvin's theorem
will remain
valid for singular solutions of the Euler equations. This question asumes some
importance
since the conservation of circulations was argued by \cite{Taylor:1938} to play
a key  role
in the enhanced production of dissipation in turbulent fluids, by the process
of vortex
line-stretching. Despite its plausibility, the validity of Taylor's argument is
far from clear.
It is not obvious {\it a priori} why there should not be anomalous dissipation
of the circulation
invariants, corresponding to a turbulent ``flux of circulations'' from large to
small scales.

In this paper, we examine these questions and establish a few relevant rigorous
results. In section \ref{classical} we briefly review the classical
Kelvin-Helmholtz
theorem and its role in turbulence dynamics. In section  \ref{general} we
prove a
theorem on conservation of circulations for singular solutions of
incompressible Euler
equations, analogous to that of \cite{Onsager:1949}  for conservation of
energy.
In section \ref{speculate} we discuss difficulties in formulating the Kelvin
theorem for
singular solutions, due to the breakdown in uniqueness of Lagrangian
trajectories,
and conjecture a statistical  version of circulation-conservation which may
apply.

\section{The Classical Kelvin Theorem}
\label{classical}

The velocity field $\bu(\bx,t)$ solving the incompressible Navier-Stokes
equation
\be \partial_t \bu + (\bu\bdot\grad)\bu = -\grad p + \nu\bigtriangleup \bu,
       \,\,\,\,\,\,\,\,\,\, \grad\bdot\bu=0 \lb{INS} \ee
with $\bx\in\Lambda\subset {\bf R}^d,$ for any integer $d\geq 2,$ satisfies
the Kelvin-Helmholtz theorem in the following sense: For any closed,
rectifiable loop $C\subset \Lambda $ at an initial time $t_0,$ the
circulation $\Gamma(C,t)=\oint_{C(t)}\bu(t)\bdot d\bx
$ satisfies
\be {{d}\over{dt}}\Gamma(C,t)=\nu\oint_{C(t)}\bigtriangleup\bu(t)\bdot d\bx,
\lb{kelvin} \ee
where $C(t)$ is the loop advected by the fluid velocity, at time $t.$ E.g., see
\cite{Saffman:1992}, section \S 1.6, for the standard derivation. It is worth
observing that the Kelvin theorem for all loops $C$ is formally equivalent to
the Navier-Stokes equation (\ref{INS}). Indeed, if $\bu(\bx,t)$ is a smooth
spacetime velocity field, diverence-free at all times $t$, then equation
(\ref{kelvin})  implies that
\be \oint_{C}\left[D_t\bu(t)-\nu\bigtriangleup\bu(t)\right]\bdot d\bx=0
\lb{K-INS}
\ee
for all loops $C$ at every time $t.$ Here
$D_t\bu=\partial_t\bu+(\bu\bdot\grad)\bu$
is the Lagrangian time-derivative and the equation (\ref{K-INS}) is derived by
applying (\ref{kelvin}) to the pre-image of the loop $C$ at initial time $t_0.$
By Stokes theorem, equation (\ref{K-INS}) can hold for all loops $C\subset
\Lambda$
if and only if there exists a pressure-field $p(\bx,t)$ such that the
Navier-Stokes
equation (\ref{INS}) holds locally and also globally, if the domain $\Lambda$
is simply connected.

In the inviscid limit $\nu\rightarrow 0,$ the circulation is formally conserved
for any initial loop $C.$ The fluid equations in this limit, the incompressible
Euler equations, are the equations of motion of a classical Hamiltonian
system. They can be derived by the Hamilton-Maupertuis principle from the
action functional
\be S[\bx] = {{1}\over{2}} \int_{t_0}^{t_f} dt \int_\Lambda  d\ba \,\,
|\dot{\bx}(\ba,t)|^2 \lb{action} \ee
with the pressure field $p(\bx,t)$ a Lagrange multiplier to enforce the
incompressibility constraint. Here $\bx(\ba,t)$ is the Lagrangian flow map
which satisfies $\dot{\bx}(\ba,t)=\bu(\bx(\ba,t),t)$ with initial condition
$\bx(\ba,t_0)
=\ba.$ See \cite{Salmon:1988} for a review. This variational principle yields
the fluid equations in a Lagrangian formulation, as $\ddot{\bx}(\ba,t)=-\grad
p(\bx(\ba,t),t).$ The Eulerian formulation (\ref{INS}) (with $\nu=0$) is
obtained
by performing variations in the inverse map $\ba(\bx,t),$ or ``back-to-labels
map'', with fixed particle positions $\bx.$ This Hamiltonian system has
an infinite-dimensional gauge symmetry group consisting of all
volume-preserving
diffeomorphisms of $\Lambda,$ which corresponds to all smooth choices
of initial fluid particle labels. In this framework, the conservation of the
circulations
for all closed loops $C$ emerges as a consequence of  Noether's theorem for
the particle-relabelling symmetry. See \cite{Salmon:1988}, Section 4.

The Kelvin theorem has many profound consequences for the dynamics
of incompressible fluids. We just note here a well-known deduction for
three-dimensional turbulence by G. I. \cite{Taylor:1938}.  It is reasonable
to assume that vortex lines---or, for that matter, any material lines---will
tend to elongate under chaotic advection by a turbulent velocity field.
Incompressibility requires that the cross-sectional area of a vortex tube
formed by such lines will shrink with time. But, in that case, the
Kelvin-Helmholtz
theorem implies that the vorticity magnitude of the tube must grow.
\cite{Taylor:1938}
observed that this process of vortex line-stretching provides an intrinsic
mechanism for amplification of the net viscous dissipation $\nu\int_\Lambda
d\bx |\bomega|^2,$ where $\bomega=\grad\btimes\bu.$ More recently,
regularizations
of the Navier-Stokes equation have been proposed as model equations for
large-scale turbulence by \cite{Holm et al.:1998,Foias et al:2001}, motivated
by requiring that a Kelvin circulation theorem be preserved.

\section{A Generalized Theorem on Conservation of Circulation}
\label{general}

This line of reasoning of  \cite{Taylor:1938} is quite delicate. It assumes
certain
properties (material advection of vortex lines and conservation of
circulations)
that can hold strictly only in the $\nu\rightarrow 0$ limit. However,
conclusions are
drawn about the limiting behavior of the energy dissipation rate, which
directly
involves the kinematic viscosity $\nu$!  Furthermore, as noted by
\cite{Onsager:1949},
the solutions of the Navier-Stokes equation are not expected to remain smooth
in the inviscid limit. Thus, the  righthand side of equation (\ref{kelvin})
does not
necessarily vanish as  $\nu\rightarrow 0$. Physically, there may be a
dissipative
anomaly for the conservation of circulations. If so, then the validity of
Taylor's
vortex-stretching mechanism for turbulent energy dissipation is open to serious
question.

The formulation of turbulent conservation of circulation by a zero-viscosity
limit
is physically natural, but not the most convenient either for numerical tests
or for rigorous
mathematical analysis\footnote{For example, the Kelvin theorem in the form of
equation (\ref{kelvin}) is not proved to be valid for the global solutions
of the Navier-Stokes equation (\ref{INS}) constructed by \cite{Leray:1934}. The
difficulty
here is that the Leray regularization of (\ref{INS}) does not preserve the
Kelvin theorem,
while alternative regularizations which do, such as that of \cite{Foias et
al:2001},  in
turn modify the energy balance. See \cite{Constantin:2003}.}. We shall instead
consider
directly the singular  (or distributional) solutions $\bu\in
L^2([0,T],\Lambda)$ of the
incompressible Euler equations, with $\nu=0.$ Let $\ol{\bu}_\ell=G_\ell*\bu$
denote
the low-pass filtered velocity at length-scale $\ell,$ where
$G_\ell(\br)=\ell^{-d}G(\br/\ell)$
is a smooth filter kernel. Then $\ol{\bu}_\ell$ satisfies the following
equation
(in the sense of distributions in time):
\be \partial_t\ol{\bu}_\ell+(\ol{\bu}_\ell\bdot \grad)\ol{\bu}_\ell
   = -\grad\ol{p}_\ell+\bF_\ell
   , \lb{F-euler} \ee
where $\ol{p}_\ell$ is the filtered pressure and where $\bF_\ell=
-\grad\bdot\btau_\ell$ is the subgrid force, i.e. minus the divergence of the
stress-tensor $\btau_\ell=\ol{(\bu\,\bu)}_\ell-\ol{\bu}_\ell\ol{\bu}_\ell.$
Let us choose a rectifiable closed loop $C$ in space. We define
 $\ol{C}_\ell(t)$ as the loop $C$ advected by the filtered velocity
 $\ol{\bu}_\ell$. This definition makes sense, since the filtered velocity
 $\ol{\bu}_\ell$ is Lipschitz in space, and corresponding flow maps exist and
 are unique (\cite{DiPernaLions:1989}).  We define a ``large-scale
circulation''
with initial loop $C$ as the line-integral
$ \ol{\Gamma}_\ell(C,t) = \oint_{\ol{C}_\ell(t)}
                               \ol{\bu}_\ell(t)\cdot d\bx. $
The same calculation that establishes the Kelvin circulation theorem
for smooth solutions of Euler equations gives that
\be
\ol{\Gamma}_\ell(C,t)-\ol{\Gamma}_\ell(C,t_0)= \int_{t_0}^t d\tau\,
  \oint_{\ol{C}_\ell(\tau)}\bF_\ell(\tau)\cdot d\bx.
\lb{F-kelvin} \ee
Thus, the line-integral of $\bF_\ell$ on the RHS represents a ``flux''
to subgrid modes at length-scales $<\ell$  of circulation on the loop
$\ol{C}_\ell(\tau)$. This motivates the definition, for any loop $C$
and filter length $\ell,$
\be K_\ell(C,t)= -\oint_{C(t)} \bF_\ell(t)  \cdot d\bx    \lb{circ-flux} \ee
so that (in generalized sense) $(d/dt)\ol{\Gamma}_\ell(C,t)=-K_\ell(C,t).$

We now prove the following:

{\bf Theorem:} {\it Let} $\zeta_p$ {\it be the pth-order scaling
exponent of the velocity, in the sense that it is the maximal value
such that }
$$ {{1}\over{|\Lambda|}}\int_\Lambda d^d\bx |\delta\bu(\br;\bx)|^p =
O(|\br|^{\zeta_p}),$$
{\it for all} $|\br|\leq r_0,$ {\it where} $\delta\bu(\br;\bx)=\bu(\bx+\br)
-\bu(\bx).$ {\it Then for any smooth loop} $C\subset \Lambda$
$$ K_\ell(C)= -\oint_{C} \bF_\ell \cdot d\bx $$
{\it satisfies} $ \lim_{\ell\rightarrow 0}K_\ell(C)= 0$ {\it if }
$\zeta_p>(d-1)+(p/2)$ {\it for any} $p\geq 2.$

The special case of this result for $p=\infty$ states that the  ``circulation
flux'' will go to zero as $\ell\rightarrow 0$ if the smallest velocity
H\"{o}lder
exponent $h_{\rm min}$ is $>1/2$. This is an exact analogue of the result
of \cite{Onsager:1949} for vanishing of energy flux when $h_{\rm min}>1/3.$
One can see that it is even easier for circulation-conservation to be anomalous
than for energy-conservation.


{\it Proof:} Our argument is close to that given by \cite{Constantin et
al.:1994}
for the Onsager theorem. The following identity for the subgrid force is easily
verified:
\begin{eqnarray*}
f_i(\bx) & = & \int d^d\br (\partial_j G)(\br) \,\delta u_i(\br;\bx)
    \delta u_j(\br;\bx) \cr
& &   \,\,\,\,\,\,\,\,\,\,\,\,\,\,\,\,\,\,\,\,\,\,\,\,\,\,\,\,
   -\int d^d\br (\partial_j G)(\br) \,\delta u_i(\br;\bx)
        \int d^d\br' \,G(\br') \,\delta u_j(\br';\bx)
\end{eqnarray*}
We omit here all subscripts $\ell$ for convenience.  By this identity,
\begin{eqnarray*}
\oint_C \bF \cdot d\bx  & =  & \int d^d\br (\partial_j G)(\br) \,
\left[\oint_C \delta u_i(\br)\delta u_j(\br) dx_i\right] \cr
 & & \,\,\,\,\,\,\,\,\,\,\,\,\,\,\,\,
   -\int d^d\br (\partial_j G)(\br)  \int d^d\br' \,G(\br') \left[
   \oint_C \delta u_i(\br)\delta u_j(\br') dx_i\right]
 \end{eqnarray*}
Thus,
\begin{eqnarray*}
\left|\oint_C \bF \cdot d\bx\right|  & \leq   & \int d^d\br
|\grad G(\br)| \,
\left[\oint_C |\delta \bu(\br)|^2 ds \right] \cr
&  & \,\,\,\,\,\,\,\,\,\,\,\,\,\,\,\,
   + \int d^d\br |\grad G(\br)|  \int d^d\br' \,G(\br') \left[
   \oint_C |\delta \bu(\br)|\,|\delta \bu(\br')| ds \right]
\end{eqnarray*}
where $s$ denotes arclength along the curve $C.$
By normalization $\int d^d\br'\,G(\br')=1$ and the
inequality $ |\delta \bu(\br)|\,|\delta \bu(\br')| \leq
{{1}\over{2}}\left[ |\delta \bu(\br)|^2+|\delta \bu(\br')|^2\right] ,$
this becomes
\begin{eqnarray}
\left|\oint_C \bF \cdot d\bx\right|  & \leq   &
{{3}\over{2}}\int d^d\br |\grad G(\br)| \,
\left[\oint_C |\delta \bu(\br)|^2 ds \right] \cr
&  & \,\,\,\,\,\,\,\,\,\,\,\,\,\,\,\,\,
   + {{1}\over{2}}\int d^d\br |\grad G(\br)|  \int d^d\br' \,G(\br') \left[
   \oint_C |\delta \bu(\br')|^2 ds \right]
\lb{main-ineq} \end{eqnarray}
We now use the H\"{o}lder inequality to derive the bound
\be \oint_C |\delta \bu(\br)|^2 ds \leq [L(C)]^{(p-2)/p}
                       \left(\oint_C |\delta \bu(\br)|^p ds\right)^{2/p}
\lb{hoelder-ineq} \ee
for any $p\geq 2, $ where $L(C)$ is the length of the curve $C.$
The condition on the scaling exponent in the statement of the
theorem can rephrased as the condition that $\bu$ belong to the
Besov space $B_{p,\infty}^{\sigma_p}(\Lambda)$ with $\sigma_p
=\zeta_p/p.$ (More properly, we should replace $\sigma_p$
by $\sigma_p-\epsilon$ for any small $\epsilon>0.$) Standard
trace theorems then imply that the restriction of $\bu$ to the
submanifold $C$ of codimension $d-1$ must satisfy $\bu |_C\in
B_{p,\infty}^{\sigma_p-(d-1)/p}(C).$ See \cite{Triebel:1983},
Theorem 2.7.2. Together with inequality (\ref{hoelder-ineq}),
 this implies that
$ \oint_C |\delta \bu(\br)|^2 ds = O\left(|\br|^{2[\zeta_p-(d-1)]/p}\right)
$\footnote{Only the case $p=\infty$ rigorously
follows from standard trace theorems. The problem is that the intrinsic
Besov space norms on the submanifold $C$ measure only the increments between
points both on $C$. However, existing trace theorems imply that every
element $f\in B_{p,\infty}^{\sigma_p'}(C),\,\,\sigma_p'=\sigma_p-(d-1)/p$
is the restriction to $C$ of some element $\tilde{f}\in
B_{p,\infty}^{\sigma_p}(\Lambda).$ The result we need follows
if the semi-norm
$ |\!|\!|f|\!|\!|_{\sigma_p'} =
   \sup_{|\br|\leq \rho}{{1}\over{|\br|^{\sigma_p'}}}
   \left[\int_C ds |\delta\tilde{f}(\br)|^p\right]^{1/p}$
is equivalent to the standard Besov semi-norm on $B_{p,\infty}^{\sigma_p'}(C)$.
Here $\br$ ranges over a ball of radius $\rho$ inside $\Lambda$%
}.
If this bound is substituted into estimate (\ref{main-ineq}) for
circulation-flux,  it yields
$$ \left|\oint_C \bF \cdot d\bx\right|
                                =O\left(\ell^{2[\zeta_p-(d-1)]/p-1}\right).
$$
We see that the latter goes to zero as $\ell\rightarrow 0,$
if $\zeta_p>p/2+(d-1).$ $\,\,\,\,\Box$

As an application, consider a velocity field that is Lipschitz
regular, so that $\zeta_p=p$ for all $p\geq 1.$
In that case, it suffices to take $p>2(d-1)$ in order to show that
the circulation is conserved (in the sense that the flux vanishes
for $\ell\rightarrow 0.$) This result applies to the 2D enstrophy
cascade, since it is expected there that $\zeta_p=p,$ with only
logarithmic corrections, for all $p\geq 2.$ See \cite{Eyink:2000}.
Thus, we expect that the Kelvin theorem holds in a strong sense
---for individual realizations---in the 2D enstrophy cascade.
However, in the 3D energy cascade the conditions of the theorem
are not expected to be satisfied.

\section{The Role of Kelvin's Theorem for Singular Solutions}
\label{speculate}

Even assuming that the assumptions of our theorem are met, there
are additional difficulties in justifying constancy of the circulation
invariants. Vanishing of the circulation flux for loops of finite length, as
established in our theorem, is not sufficient. In the first place, the material
loop $\ol{C}_\ell(t)$ is not expected to remain rectifiable as $\ell\rightarrow
0$,
but instead to become a fractal curve $C(t)$ with Hausdorff dimension
$>1$ for any positive time $t$ (\cite{SreenivasanMeneveau:1986}). Thus,
we cannot immediately infer that the RHS of equation (\ref{F-kelvin}) vanishes
as $\ell\rightarrow 0,$ nor even make sense of the contour integral in that
limit.
A possible approach here is to transform the RHS to label-space, as
$\oint_C \bF_\ell(\ol{\bx}_\ell(\tau),\tau)\cdot d\ol{\bx}_\ell(\tau)$
where the map satisfies $\dot{\ol{\bx}}_\ell(\ba,\tau)=\ol{\bu}_\ell
(\ol{\bx}_\ell(\ba,\tau),\tau).$ This can make sense as a Stieltjes
integral on the loop $C$ in label-space for H\"{o}lder continuous
maps (e.g. see \cite{Young:1936}),\cite{Zahle:1998}).

However, there is a much more serious problem in formulating Kelvin's
theorem for singular Euler solutions: it is not clear that material loops
exist! Recent work on an idealized turbulence problem---the Kraichnan
model of random advection---has shown that Lagrangian particle
trajectories $\bx(t),\,\bx'(t)$ can explosively separate even when
$\bx_0=\bx'_0$ initially, if the advecting velocity field is only
H\"{o}lder continuous and not Lipschitz. See \cite{Bernardetal:1998}.
Mathematically, this is a consequence
of the non-uniqueness of solutions to the initial-value problem, while,
physically, it corresponds to the two-particle turbulent diffusion of
\cite{Richardson:1926}. \cite{LeJanRaimond:2002,LeJanRaimond:2004}
have rigorously proved that there is a random process of Lagrangian
particle paths $\bx(t)$ in the Kraichnan model for a fixed realization of the
advecting velocity and a fixed initial particle position. This phenomenon
has been termed {\it spontaneous stochasticity} (\cite{Chaves et al:2003}).
A similar notion of  ``generalized flow'' was proposed by \cite{Brenier:1989}
for the problem of minimizing the action (\ref{action}). In his formulation,
the
action  is generalized to a functional $S[P]={{1}\over{2}} \int P(d\bx)
\int_{t_0}^{t_f} dt
\,\,|\dot{\bx}(t)|^2$, where $P$ is a probability measure on path-space,
and he showed that minimizers always exist in this framework. Unfortunately,
this notion does not permit one to define the concept of material lines
and surfaces for ideal flow. A more natural generalization of the classical
action would be of the form
\be S[P] = {{1}\over{2}}\int P(d\bx) \int_{t_0}^{t_f} dt \int_\Lambda  d\ba
\,\,
|\dot{\bx}(\ba,t)|^2 \lb{P-action} \ee
where $P$ is now a probability measure on time-histories of measure-preserving
maps \footnote{The Kraichnan model might also benefit from a formulation
in terms of maps. Formally, a group of Markov transition operators
$S_{t,t'}^\bu$
can be defined on spaces of functionals of maps, with a fixed realization of
the
velocity $\bu$, via a  Krylov-Veretennikov expansion:
\begin{eqnarray*}
&& S_{t,t'}^{\bu}=\sum_{n=0}^\infty (-1)^n \int_{t'}^t dt_1\int_\Lambda d\ba_1
\int_{t'}^{t_1} dt_2\int_\Lambda d\ba_2\cdots\int_{t'}^{t_{n-1}}
dt_n\int_\Lambda d\ba_n \cr
&&
e^{(t-t_1)\cL_0}\left(\bu(\bx(\ba_1),t_1)
\bdot{{\delta}\over{\delta\bx(\ba_1)}}\right)
e^{(t_1-t_2)\cL_0}\left(\bu(\bx(\ba_2),t_2)
\bdot{{\delta}\over{\delta\bx(\ba_2)}}\right)
e^{(t_2-t_3)\cL_0} \cr
&& \,\,\,\,\,\,\,\,\,\,\,\,\,\,\,\,\,\,\,\,\,\,\,\,\,\,\,\,\,\,\,\,\,\,\,\,\,\,
\,\,\,\,\,\,\,\,\,\,\,\,\,\,\,\,\,\,
\,\,\,\,\,\,\,\,\,\,\,\,\,\, \cdots
e^{(t_{n-1}-t_n)\cL_0}\left(\bu(\bx(\ba_n),t_n)
\bdot{{\delta}\over{\delta\bx(\ba_n)}}\right)
e^{(t_n-t')\cL_0}.
\end{eqnarray*}
Cf. \cite{LeJanRaimond:2002,LeJanRaimond:2004}. Here the time-integrals should
be defined in the Ito sense with respect to the white-noise velocity field
$\bu(\bx,t)$
and $\cL_0$ is formally the infinitesimal generator of a diffusion process on
the
space of maps, given by $\cL_0= {{1}\over{2}}\int_\Lambda d\ba\int_\Lambda
d\ba'
D_{ij}(\bx(\ba)-\bx(\ba')){{\delta^2}\over{\delta x_i(\ba)\delta x_j(\ba')}}$.
The Gaussian
random velocity has covariance $\langle
u_i(\bx,t)u_j(\bx',t')\rangle=D_{ij}(\bx-\bx')
\delta(t-t').$ It would be very interesting to give rigorous meaning to this
expansion,
especially for the case where the advecting velocity is only H\"{o}lder
continuous
but not Lipschitz in space}. For any realization of such a random process and
for any
initial curve $C$ the advected object $C(t)=\bx(C,t)$ is well-defined and
remains
a (random) curve for all time $t,$ if the maps are continuous in space.

Let us assume for the moment that the (very nontrivial) problem can be
solved to construct such a generalized flow $\bx(\ba,t),$ or stochastic process
in the space  of volume-preserving maps,  which is hopefully a.s. H\"{o}lder
continuous in space so that material loops $C(t)$ exist as random, fractal
curves.
We would like to present some plausibility arguments in favor of the conjecture
that circulations shall be conserved in a statistical sense.  More precisely,
we
expect that the circulations $\Gamma(C,t)$ for any initial smooth loop $C$
shall be martingales of the generalized flow:
\be
E\left[\Gamma(C,t)|\Gamma(C,\tau),\tau<t'\right]=\Gamma(C,t'),\,\,\,\mbox{for
$t>t'.$}
\lb{martingale} \ee
Here $E[\cdot]$ denotes the expectation over the ensemble of random Lagrangian
paths
and we have conditioned on the past circulation history
$\{\Gamma(C,\tau),\tau<t'\}. $
Heuristically,
\be  (d/dt)E\left[\Gamma(C,t)|\Gamma(C,\tau),\tau<t'\right]=
       -\lim_{\ell\rightarrow
0}E\left[K_\ell(C,t)|\Gamma(C,\tau),\tau<t'\right]. \lb{dot-circ} \ee
Note that the conditioning event involves scales of the order of the radius
of gyration of the loops $C(\tau),\,\,\tau<t',$ while the circulation-flux
involves
velocity-increments over separation lengths $\ell\rightarrow 0.$ Therefore,
we expect that Kolmogorov's idea of small-scale homogeneity (and isotropy)
will apply. Note, however, that the homogeneous average of the subgrid
force $\bF_\ell$ is zero, because it is the divergence of the stress tensor.
{}From another point of view, the subgrid force will become increasingly
irregular for $\ell\ll R(t)$ (the radius of the loop $C(t)$) and the sign
of the integrand $\bF_\ell(\ol{\bx}_\ell(s,t))\bdot\ol{\bx}'_\ell(s,t)$ will
oscillate
more rapidly as a function of the arclength $s.$ Thus, cancellations will
occur. For these reasons, we expect that the limit on the RHS of
(\ref{dot-circ}) shall vanish, implying (\ref{martingale}).  Another formal
argument can be given by applying the Noether theorem to the generalized
action (\ref{P-action}) and using the fact that a global minimizer must
also minimize the action for the time segment $[t',t_f].$  On the other hand,
based upon our earlier theorem, it is very unlikely that circulation-flux will
vanish as $\ell\rightarrow 0$ in every realization, without any averaging.

In this section we have clearly indulged in some speculative thinking,
but we hopefully have also succeeded in outlining the various difficulties
in properly formulating Kelvin's theorem for turbulent solutions of the Euler
equations.
Our own view is that the \cite{Taylor:1938} mechanism of vortex line-stretching
is the underlying cause of enhanced dissipation in three-dimensional turbulence
asymptotically at high Reynolds numbers. However, much work remains to
elucidate
the details of the subtle dynamics involved.


\end{document}